\documentclass{PoS}

\usepackage{mathpazo}
\usepackage{amsmath,amssymb,amsfonts,amsthm}
\usepackage{enumerate} 
\usepackage{multirow} \usepackage{rotating}

\def\f{\frac}

\newcommand{\C}{\mathbb{C}}

\newcommand{\be}{\nopagebreak[3]\begin{equation}}
\newcommand{\ee}{\end{equation}}
\newcommand{\ba}{\nopagebreak[3]\begin{eqnarray}}
\newcommand{\ea}{\end{eqnarray}}
\newcommand{\nn}{\nonumber \\ }

\newcommand{\beq}{\begin{eqnarray}}
\newcommand{\eeq}{\end{eqnarray}}
\newcommand{\bea}{\begin{eqnarray}}
\newcommand{\eea}{\end{eqnarray}}

\newcommand{\N}{{\mathbb N}}
\newcommand{\R}{{\mathbb R}}

\newcommand{\pP}{{\mathbb P}}

\newcommand{\cL}{{\mathcal L}}
\newcommand{\cH}{{\mathcal H}}

\newcommand{\cM}{{\mathcal M}}

\newcommand{\cC}{{\mathcal C}}

\newcommand{\SU}{\mathrm{SU}}

\newcommand{\U}{\mathrm{U}}

\newcommand{\vJ}{\vec{J}}

\newcommand{\id}{\mathbb{I}}

\newcommand{\su}{{\mathfrak su}}
\renewcommand{\sl}{{\mathfrak sl}}

\renewcommand{\u}{{\mathfrak u}}
\newcommand{\la}{\langle}
\newcommand{\ra}{\rangle}

\newcommand{\tr}{{\mathrm Tr}}

\def\eps{\epsilon}

\newcommand{\bz}{\overline{z}}

\newcommand{\Ref}[1]{(\ref{#1})}

\def\pp{\partial}

\def\bz{\bar{z}}
\def\bQ{\bar{Q}}
\def\inv{{\textrm{Inv}}}
\def\cHNJ{\cH_N^{(J)}}
\def\cHN{\cH_N}
\def\tU{{{}^t U}}
\def\vcC{\vec{{\cal C}}}
\def\cHQJ{\cH^{(Q)}_J}

\def\wM{\widehat{M}}
\def\wQ{\widehat{Q}}
\def\wbQ{\widehat{\bar{Q}}}
\def\wcC{\widehat{\cC}}

\newcommand{\mat}[2]{\left(\begin{array}{#1}#2
\end{array}\right)}

\title{$\U(N)$ and holomorphic methods for LQG and Spin Foams}

\ShortTitle{$\U(N)$ and holomorphic methods for LQG and Spin Foams}

\author{Enrique F. Borja\\
        Institute for Theoretical Physics III, University of
Erlangen-N\"{u}rnberg,\\ Staudtstra{\ss}e 7, D-91058 Erlangen (Germany).\\
Departamento de F\'{\i}sica Te\'{o}rica and IFIC, Centro Mixto Universidad de
Valencia-CSIC. Facultad de F\'{\i}sica, Universidad de Valencia,
Burjassot-46100, Valencia (Spain).\\
        E-mail: \email{efborja@theorie3.physik.uni-erlangen.de}}

\author{Jacobo Diaz-Polo\\
        Department of Physics and Astronomy, Louisiana State University\\
        Baton Rouge, LA, 70803-4001.\\
        E-mail: \email{jacobo@phys.lsu.edu}}

\author{I\~naki Garay\\
        Institute for Theoretical Physics III, University of
Erlangen-N\"{u}rnberg,\\ Staudtstra{\ss}e 7, D-91058 Erlangen (Germany).\\
        E-mail: \email{igael@theorie3.physik.uni-erlangen.de}}

\abstract{ The $\U(N)$ framework and the spinor representation for
loop quantum gravity are two new points of view that can help us
deal with the most fundamental problems of the theory. Here, we
review the detailed construction of the $\U(N)$ framework explaining
how one can endow the Hilbert space of $N$-leg intertwiners
with a Fock structure. We then give a description of the
classical phase space corresponding to this system in terms of the
spinors, and we will study its quantization using holomorphic
techniques. We take special care in constructing the usual
holonomy operators of LQG in terms of spinors, and in the
description of the Hilbert space of LQG with the different
polarization given by these spinors.}

\FullConference{3rd Quantum Gravity and Quantum Geometry School \\
         February 28 - March 13, 2011\\
         Zakopane, Poland}

\begin{document}


\section{Introduction}

Over the last few years, a new set of tools for approaching some of
the fundamental open problems in loop quantum gravity (LQG) has been
developed. Since the identification in
\cite{Girelli:2005ii,Freidel:2009ck} of a characteristic $\U(N)$
symmetry in the Hilbert space of $\SU(2)$ intertwiners, a new
framework has emerged, producing some exciting results, but most
importantly, opening a brand new avenue to look at fundamental
issues like the dynamics of the theory or the identification of
symmetries at a purely quantum level. The $\U(N)$ framework can also
be derived as the holomorphic quantization of a classical spinor
system, providing, among other interesting insights, a novel way to
study the semiclassical limit of the theory. In this article we
present a short but self-contained pedagogical introduction to the
core formulation and techniques that constitute the new $\U(N)$
framework for loop quantum gravity, including the related spinor
formulation and holomorphic quantization methods.

The organization of the paper is as follows. In section 2 we present
the so-called $\U(N)$ framework for loop quantum gravity. We introduce
a spinorial representation in section 3 and proceed to its quantization 
using holomorphic methods in section 4, providing a full analogy with the
$\U(N)$ framework. In section 5 we discuss an action principle
for the classical spinor system. Finally, in section 6 we
present a short review of the ``state of the art'' on the
holomorphic methods based on the work by Livine and Tambornino
\cite{Livine:2011gp}, giving strong formal consistency to the
techniques presented here. We conclude summarizing the
main features of the presented framework.


\section{The $\U(N)$ framework for intertwiners}

Our starting point to introduce the $\U(N)$ framework will be to
study the structure of the Hilbert space of $\SU(2)$ intertwiners,
the building blocks of spin network states. This is, given a set of
$\SU(2)$ representations (spins), the space of invariant tensors
\begin{equation}
\cH_{j_1,j_2,\ldots}\,\equiv\, \textrm{Inv}[V^{j_1}\otimes V^{j_2}\otimes\ldots]\,,\nonumber
\end{equation}
where $V^{j_i}$ are the irreducible representation spaces associated
to the spins $j_1, j_2,\ldots$. In loop quantum gravity these
intertwiners can be dually regarded as a region of space with a
(topologically) spherical boundary punctured by the intertwiner's
legs.

In particular, let us consider the space of $N$-valent intertwiners
and fixed total spin $J=\sum_i j_i$ (that can be thought of as the
total area of the surface enclosing the intertwiner):
\begin{equation}
\cH_N^{(J)}\,\equiv\,\bigoplus_{\sum_i
j_i=J}\cH_{j_1,..,j_N}.\nonumber
\end{equation}

As shown in \cite{Freidel:2009ck}, intertwiner spaces $\cH_N^{(J)}$
carry irreducible representations of $\U(N)$ and the full space
$\cH_N$ can be endowed with a Fock space structure with creation and
annihilation operators compatible with the $\U(N)$ action
\cite{Freidel:2010tt}. This structure is at the foundation of the
$\U(N)$ techniques, and we will review its basic construction in
what follows.

We start by introducing the well-known Schwinger representation of
the $\su(2)$ algebra. This representation describes the generators
of $\su(2)$ in terms of a pair of uncoupled harmonic oscillators. In
our case, we introduce $2N$ oscillators --a pair for each leg of the
intertwiner-- with creation operators $a_i,b_i$, $i$ running from
$1$ to $N$:
\begin{equation}
[a_i,a^\dag_j]=[b_i,b^\dag_j]=\delta_{ij}\,,\qquad
[a_i,b_j]=0.\nonumber
\end{equation}

The local $\su(2)$ generators at each
leg of the intertwiner can be expressed then as quadratic operators:
\begin{equation}
J^z_i=\f12(a^\dag_i a_i-b^\dag_ib_i),\qquad J^+_i=a^\dag_i
b_i,\qquad J^-_i=a_i b^\dag_i,\qquad E_i=(a^\dag_i a_i+b^\dag_ib_i).
\end{equation}
As expected, the $J_i$'s constructed this way satisfy the
standard commutation relations of the $\su(2)$ algebra while the
total energy $E_i$ is a Casimir operator:
\begin{equation}
[J^z_i,J^\pm_i]=\pm J^\pm_i,\qquad [J^+_i,J^-_i]=2J^z_i,\qquad
[E_i,\vec{J}_i]=0.\nonumber
\end{equation}
The operator $E_i$ is the total energy carried by the pair of
oscillators $a_i,b_i$ and its eigenvalue is twice the spin $2j_i$ of the
corresponding $\SU(2)$ representation. We can express
the standard $\SU(2)$ Casimir operator in terms of this energy as:
\begin{equation}
\vJ_i^2\,=\, \f{E_i}2\left(\f{E_i}2+1\right)\,=\,
\f{E_i}4\left({E_i}+2\right).\nonumber
\end{equation}

As it is well-known, in the context of LQG the spin $j_i$ is related
to the area associated with the $i$-th leg of the intertwiner. In
this particular framework, the most natural choice for the
regularization of the area operator is such that the spectrum is
given directly by the Casimir $E_i/2$ (the spin $j_i$); we will
consider that case in this paper.

The key observation now is that one can use these harmonic oscillators to construct operators acting on the
Hilbert space of intertwiners, i.e., operators that are
invariant under global $\SU(2)$ transformations generated by
$\vJ\,\equiv\,\sum_i \vJ_i$. These constitute the starting point of the $\U(N)$ formalism, and they are quadratic
invariant operators acting on pairs of (possibly equal) legs $i,j$
\cite{Girelli:2005ii,Freidel:2009ck}:
\beq
&&E_{ij}=a^\dag_ia_j+b^\dag_ib_j, \qquad E_{ij}^\dag=E_{ji},\nn
&&F_{ij}=(a_i b_j - a_j b_i),\qquad F_{ji}=-F_{ij}.\nonumber
\eeq
The operators $E,F,F^\dag$ form a closed algebra:
\bea
\label{commEF} {[}E_{ij},E_{kl}]&=&
\delta_{jk}E_{il}-\delta_{il}E_{kj},\\
{[}E_{ij},F_{kl}] &=& \delta_{il}F_{jk}-\delta_{ik}F_{jl},\qquad
{[}E_{ij},F_{kl}^{\dagger}] =
\delta_{jk}F_{il}^{\dagger}-\delta_{jl}F_{ik}^{\dagger}, \nn
{[}F_{ij},F^{\dagger}_{kl}] &=& \delta_{ik}E_{lj}-\delta_{il}E_{kj}
-\delta_{jk}E_{li}+\delta_{jl}E_{ki}
+2(\delta_{ik}\delta_{jl}-\delta_{il}\delta_{jk}), \nn
{[}F_{ij},F_{kl}] &=& 0,\qquad {[}
F_{ij}^{\dagger},F_{kl}^{\dagger}] =0.\nonumber
\eea

We can see that commutators of $E_{ij}$ operators have the structure
of a $\u(N)$-algebra (which motivates the name of the $\U(N)$
framework). The diagonal operators are precisely the operators
giving the energy on each leg, $E_{ii}=E_i$. Then the value of the
total energy $E\,\equiv \sum_i E_i$ gives twice the sum of all spins
$2\times\sum_i j_i$, i.e. twice the total area.

The $E_{ij}$-operators change the energy/area carried by each leg,
while still conserving the total energy, while the operators
$F_{ij}$ (resp. $F^\dag_{ij}$) decrease (resp. increase) the
total area $E$ by $2$:
\begin{equation}
[E,E_{ij}]=0,\qquad [E,F_{ij}]=-2F_{ij},\quad
[E,F^\dag_{ij}]=+2F^\dag_{ij}.\nonumber
\end{equation}
This suggests that we can decompose the Hilbert space of $N$-valent
intertwiners into subspaces of constant area:
\begin{equation}
\cH_N=\bigoplus_{\{j_i\}} \inv\left[\otimes_{i=1}^NV^{j_i}\right]
=\bigoplus_{J\in\N}\bigoplus_{\sum_ij_i=J}
\inv\left[\otimes_{i=1}^NV^{j_i}\right] =\bigoplus_J
\cH_N^{(J)},\nonumber
\end{equation}
where as before $V^{j_i}$ denote the Hilbert space of the irreducible
$\SU(2)$-representation of spin $j_i$, spanned by the states  of the
oscillators $a_i,b_i$ with fixed total energy $E_i=2j_i$.

In \cite{Freidel:2009ck}, the structure of these subspaces $\cHNJ$
of $N$-valent intertwiners with fixed total area $J$ was studied,
identifying the irreducible representations of $\U(N)$, generated by
the $E_{ij}$ operators, that they naturally carry. Then the
operators $E_{ij}$ allow to navigate from state to state within each
subspace $\cHNJ$. On the other hand, operators
$F_{ij},\,F^\dag_{ij}$ allow to go from one subspace $\cHNJ$ to the
next $\cHN^{(J\pm 1)}$, thus endowing the full space of $N$-valent
intertwiners $\cH_N$ with a Fock space structure, with creation
operators $F^\dag_{ij}$ and annihilation operators $F_{ij}$.

Finally, it was also found that the whole set of operators
$E_{ij},F_{ij},F^\dag_{ij}$ satisfy the following quadratic constraints
\cite{Borja:2010gn}:
\beq
\label{quadratic1}
\forall i,j,\quad &&\sum_k E_{ik}E_{kj}=E_{ij} \left(\f
E2+N-2\right),\label{constraint0}\\
&&\label{quadratic2}
\sum_k F^\dagger_{ik}E_{jk} =
F^\dagger_{ij}\, \frac{E}{2}, \qquad\qquad\quad
\sum_k E_{jk} F^\dagger_{ik} = F^\dagger_{ij}\left(\frac{E}{2}+N-1\right),\label{constraint1}\\
&&\label{quadratic3}
\sum_k E_{kj}F_{ik} = F_{ij}\, \left(\frac{E}{2}-1\right), \qquad
\sum_k F_{ik} E_{kj}  = F_{ij}\left(\frac{E}{2}+N-2\right),\label{constraint2}\\
&&\label{quadratic4}
\sum_k F^\dagger_{ik}F_{kj} = E_{ij}
\left(\frac{E}{2}+1\right),\qquad \sum_k F_{kj}F^\dag_{ik} =
(E_{ij}+2\delta_{ij})
\left(\frac{E}{2}+N-1\right)\,.\label{constraint3}
\eeq
As noticed in \cite{Borja:2010gn} and further extended in
\cite{Borja:2010rc}, these relations have the structure of
constraints on the multiplication of two matrices $E_{ij}$ and
$F_{ij}$. This is one of the main hints that will lead to the
derivation of the $\U(N)$ framework as a quantization of a classical
matrix model, as we are going to see in following sections.


\section{Classical spinor formalism}

A very interesting feature of the new $\U(N)$-framework is that it
can be re-derived in terms of spinors in a rather straightforward
way \cite{Livine:2011gp,Borja:2010rc}. The operators in the
$\U(N)$-formalism can be shown to arise as the quantization of
classical spinor matrices. This connection can help understand the
geometrical meaning of spin network states in LQG, as well as
provide hints on the semi-classical limit of the full theory. There
is also a connection with the so-called ``twisted geometries''
\cite{Freidel:2010aq,Freidel:2010bw} that express the classical
phase space of loop quantum gravity on a given graph as a classical
spinor model, unravelling the relation between spin networks and
discrete geometry. This could provide new ideas on the study of spin
network dynamics through the spinfoam approach.

We are going to present the basic concepts that lead to recast the
$\U(N)$-framework in terms of spinors, showing how this is related
to standard $\SU(2)$ intertwiners in LQG. Let us start by
introducing the spinor notation that we will use
\cite{Livine:2011gp,Freidel:2010tt,Borja:2010rc,Freidel:2010bw,Dupuis:2010iq}.
Let $z$ be a spinor
$$
|z\ra=\mat{c}{z^0\\z^1}, \qquad \la z|=\mat{cc}{\bar{z}^0
&\bar{z}^1}.
$$
We can associate it to a geometrical 3-vector $\vec{X}(z)$, defined
from the projection of the $2\times 2$ matrix $|z\ra\la z|$ onto
Pauli matrices $\sigma_a$ (taken Hermitian and normalized so that
$(\sigma_a)^2=\id$):
\begin{equation} \label{vecV}
|z\ra \la z| = \f12 \left( {\la z|z\ra}\id  +
\vec{X}(z)\cdot\vec{\sigma}\right).
\end{equation}
It is straightforward to compute the norm and the components of this
vector in terms of the spinors:
\begin{equation*}
|\vec{X}(z)| = \la z|z\ra= |z^0|^2+|z^1|^2,\qquad
X^z=|z^0|^2-|z^1|^2,\quad X^x=2\,\Re\,(\bar{z}^0z^1),\quad
X^y=2\,\Im\,(\bar{z}^0z^1).
\end{equation*}
With this, the spinor $z$ is entirely determined by the
corresponding 3-vector $\vec{X}(z)$ up to a global phase. We can
give the reverse map:
\begin{equation*}
z^0=e^{i\phi}\,\sqrt{\f{|\vec{X}|+X^z}{2}},\quad
z^1=e^{i(\phi-\theta)}\,\sqrt{\f{|\vec{X}|-X^z}{2}},\quad
\tan\theta=\f{X^y}{X^x},
\end{equation*}
where $e^{i\phi}$ is an arbitrary phase.

Let us also introduce now the duality map $\varsigma$ acting on spinors:
\begin{equation*}
\varsigma\begin{pmatrix}z^0\\ z^1\end{pmatrix} \,=\,
\begin{pmatrix}-\bar{z}^1\\\bar{z}^0 \end{pmatrix},
\qquad \varsigma^{2}=-1.
\end{equation*}
This is an anti-unitary map, $\la \varsigma z| \varsigma w\ra= \la
w| z\ra=\overline{\la z| w\ra}$, and we will write the related state
as
$$
|z]\equiv \varsigma  | z\ra,\qquad [z| w]\,=\,\overline{\la z|
w\ra}.
$$
This map $\varsigma$ maps the 3-vector $\vec{X}(z)$ onto its
opposite:
\begin{equation}
|z][  z| = \f12 \left({\la z|z\ra}\id -
\vec{X}(z)\cdot\vec{\sigma}\right).\nonumber
\end{equation}

In order now to describe $N$-valent intertwiners,
 we consider $N$ spinors $z_i$ and their corresponding
3-vectors $\vec{X}(z_i)$. A standard requirement is to ask the $N$
spinors to satisfy a closure condition, i.e., that the sum of the
corresponding 3-vectors vanish, $\sum_i \vec{X}(z_i)=0$. Recalling
the definition of $\vec{X}(z_i)$, this closure condition can be
expressed in terms of $2\times 2$ matrices:
\begin{equation}
\sum_i |z_i\ra \la z_i|=A(z)\id, \qquad\textrm{with}\quad
A(z)\equiv\f12\sum_i \la z_i|z_i\ra=\f12\sum_i|\vec{X}(z_i)|.
\end{equation}
This further translates into quadratic constraints on the spinors:
\begin{equation}
\sum_i z^0_i\,\bar{z}^1_i=0,\quad \sum_i \left|z^0_i\right|^2=\sum_i
\left|z^1_i\right|^2=A(z). \label{closure}
\end{equation}
In simple terms, this means that the two components of the spinors,
$z^0_i$ and $z^1_i$, form orthogonal $N$-vectors of equal norm. In
order to simplify the notation, let us introduce the matrix elements
of the $2\times 2$ matrix $\sum_i |z_i\ra\la z_i|$~:
\begin{equation}
\cC_{ab}=\sum_i z^a_i \bz^b_i.\nonumber
\end{equation}
Then the closure constraints are written very simply:
\begin{equation}
\cC_{00}-\cC_{11}=0,\quad \cC_{01}=\cC_{10}=0.\nonumber
\end{equation}


\section{(Anti-)holomorphic quantization}

Let us construct the classical phase space in terms of spinors. We
will then proceed to its quantization, following
\cite{Livine:2011gp,Borja:2010rc}.

We start by postulating simple Poisson bracket relations for a set of
$N$ spinors:
\begin{equation}
\label{bracketz}
\{z^a_i,\bz^b_j\}\,\equiv\,i\,\delta^{ab}\delta_{ij},
\end{equation}
with all other brackets vanishing,
$\{z^a_i,z^b_j\}=\{\bz^a_i,\bz^b_j\}=0$.
They exactly reproduce the Poisson bracket structure of $2N$ uncoupled harmonic oscillators.

Our expectation now is to have the closure constraints generating global $\SU(2)$
transformations on the $N$ spinors. Let us then compute the Poisson brackets between
components of
the $\cC$-constraints~:
\beq
\label{commC}
\hspace*{-1cm}&&\{\cC_{00}-\cC_{11},\cC_{01}\}=\!-2i\cC_{01},\quad\!\!
\{\cC_{00}-\cC_{11},\cC_{10}\}=\!+2i\cC_{10},\quad\!\!
\{\cC_{10},\cC_{01}\}=i(\cC_{00}-\cC_{11}),\\
\hspace*{-1cm}&&\{\tr \cC,\cC_{00}-\cC_{11}\}=\{\tr
\cC,\cC_{01}\}=\{\tr \cC,\cC_{10}\}=0.\nonumber
\eeq
These four components $\cC_{ab}$ do indeed form a closed $\u(2)$
algebra, with the three closure conditions $\cC_{00}-\cC_{11}$,
$\cC_{01}$ and $\cC_{10}$ being generators of a $\su(2)$ subalgebra.
We will denote as $\vec{\cC}$ these three $\su(2)$-generators, with
$\cC^z\equiv\cC_{00}-\cC_{11}$ and $\cC^+=\cC_{10}$ and
$\cC^-=\cC_{01}$.
One can already guess that these three closure conditions $\vec{\cC}$ will be related to the
$SU(2)$ generators $\vJ$ at the quantum level, while the operator $\tr\,\cC$
will correspond to the total energy/area $E$.

In correspondence to operators $E$ and $F$, let us introduce
matrices $M$ and $Q$ satisfying $M=M^\dag,\, {}^tQ=-Q$ and the
classical analogs to the quadratic constraints
(\ref{quadratic1}--\ref{quadratic4}). Up to a global phase, these
matrices can be written with generality as:
\beq
&&M=\lambda\,U\Delta U^{-1},\qquad\,\, \Delta=\mat{cc|c}{1 & & \\ &1
& \\ \hline && 0_{N-2}}\,, \nn
&&Q=\lambda\,U\Delta_\eps \tU,\qquad \Delta_\eps=\mat{cc|c}{ & 1& \\
-1& & \\ \hline & &0_{N-2}}\,, \nonumber
\eeq
where $U$ is a unitary matrix $U^\dag U=\id$. Then, if we define spinors such that
\begin{equation}
z_i\,\equiv\,\mat{c}{\bar{u}_{i1}\sqrt{\lambda}\\
\bar{u}_{i2}\sqrt{\lambda}}\,,\qquad \lambda\equiv\tr M/2,
\end{equation}
$u_{ij}$ being the elements of the unitary matrix $U$, we can write the components of $M$ and $Q$
as
\begin{equation}
M_{ij}=\la z_i |z_j \ra=\overline{\la z_j |z_i \ra}, \qquad
Q_{ij}=\la z_j |z_i]=\overline{[ z_i |z_j \ra}=-\overline{[ z_j |z_i
\ra}.
\end{equation}
One can see that, in this setting, the unitarity condition on
the matrices $U$ is equivalent to the closure
conditions on the spinors.

If we now compute the Poisson brackets of the $M_{ij}$ and
$Q_{ij}$ matrix elements:
\beq
&&\{M_{ij},M_{kl}\}=i(\delta_{kj}M_{il}-\delta_{il}M_{kj}),
\label{commM}\nn
&&\{M_{ij},Q_{kl}\}=i(\delta_{jk}Q_{il}-\delta_{jl}Q_{ik}),\nn
&&\{Q_{ij},Q_{kl}\}=0,\nn
&&\{\bar{Q}_{ij},Q_{kl}\}=i(\delta_{ik}M_{lj}+\delta_{jl}M_{ki}-\delta_{jk}M_{li}-\delta_{il}M_{kj}),\nonumber
\eeq
we observe that they reproduce the expected commutators \Ref{commEF} up to the
$i$-factor. We further check that these variables commute with the
closure constraints generating the $\SU(2)$ transformations:
\begin{equation}
\{\vcC,M_{ij}\}=\{\vcC,Q_{ij}\}=0.
\end{equation}
Finally, one can also compute their commutator with $\tr \,\cC$:
\begin{equation}
\{\tr\,\cC,M_{ij}\}=0,\quad \{\tr\,\cC,Q_{ij}\}=\{\sum_k
M_{kk},Q_{ij}\}=\,+2i\,Q_{ij},\nonumber
\end{equation}
which confirms that the matrix $M$ is invariant under the full
$\U(2)$ subgroup and that $\tr\,\cC$ acts as a dilatation operator
on the $Q$ variables, or reversely that the $Q_{ij}$ acts as
creation operators for the total energy/area variable $\tr\,\cC$.

\medskip

We have already characterized the classical phase space
associated to spinors $z_i$ and matrices $M_{ij},Q_{ij}$.
Let us now proceed to its quantization. In order to do so, we
will consider Hilbert spaces $\cHQJ$ of homogeneous polynomials in
$Q_{ij}$ of degree $J$:
\begin{equation}
\cHQJ \,\equiv\, \{P \in \pP[Q_{ij}] \,|\quad P(\rho
Q_{ij})\,=\,\rho^J\,P(Q_{ij}),\,\forall\rho\in\C  \}\,.
\end{equation}
These are completely anti-holomorphic polynomials in $z_i$
(holomorphic in $\bar{z}_i$) and of order $2J$.

One can show that these Hilbert spaces $\cHQJ$ are isomorphic to the
Hilbert space $\cHNJ$  of $N$-valent intertwiners with fixed total
area $J$. To this purpose, we will construct the explicit
representation of the operators resulting from the quantization of
$M_{ij}$ and $Q_{ij}$ on $\cHQJ$ and show that they match the action
of the $\U(N)$ operators $E_{ij}$ and $F^\dag_{ij}$ described
earlier. We choose to quantize $\bz_i$ as multiplication operators
while promoting $z_i$ to derivative operators:
\begin{equation}
\widehat{\bz}_i^a \,\equiv\, {\bz}_i^a\,\times\,,\qquad
\widehat{z}_i^a \,\equiv\, \f{\pp}{\pp \bz_i^a},
\end{equation}
which satisfy the commutator $[\hat{z},\hat{\bz}]=1$ as expected
for the quantization of the classical bracket $\{z,\bz\}=i$. We can then
quantize the matrix elements $M_{ij}$ and $Q_{ij}$ and the
closure constraints following this correspondence:
\beq
\widehat{M}_{ij} &=& {\bz}_i^0\f{\pp}{\pp
\bz_j^0}+{\bz}_i^1\f{\pp}{\pp \bz_j^1}\,, \nn
\widehat{Q}_{ij} &=&
\bz_i^0\bz_j^1-\bz_i^1\bz_j^0 \,=\, Q_{ij}\,, \nn
\widehat{\bQ}_{ij}
&=& \f{\pp^2}{\pp \bz_i^0\pp \bz_j^1}-\f{\pp^2}{\pp \bz_i^1\pp
\bz_j^0}\,, \nn
\widehat{\cC}_{ab} &=&
\sum_k{\bz}_k^b\f{\pp}{\pp\bz_k^a}.\nonumber
\eeq
It is straightforward to check that the $\wcC_{ab}$ and the
$\wM_{ij}$ respectively form a $\u(2)$ and a $\u(N)$ Lie algebra, as
expected:
\begin{equation}
[\wcC_{ab},\wcC_{cd}]\,=\,
\delta_{ad}\wcC_{cb}-\delta_{cb}\wcC_{ad},\qquad
[\wM_{ij},\wM_{kl}]\,=\,
\delta_{kj}\wM_{il}-\delta_{il}\wM_{kj},\qquad
[\wcC_{ab},\wM_{ij}]\,=\,0.
\end{equation}
which amounts to multiply the Poisson bracket \ref{commC} and
\ref{commM}  by $-i$.

Let us analyze the structure resulting from this quantization. First, by checking the action of the closure constraints on
functions of the variables $Q_{ij}$~:
\beq
&&\widehat{\vcC}\,Q_{ij}\,=\, 0 ,\qquad
\widehat{(\tr\,\cC)}\,Q_{ij}\,=\, 2Q_{ij},\nn \forall
P\in\cHQJ=\pP_J[Q_{ij}],\qquad
&&\widehat{\vcC}\,P(Q_{ij})\,=\,0,\qquad
\widehat{(\tr\,\cC)}\,P(Q_{ij})\,=\, 2J\,P(Q_{ij}),\nonumber
\eeq
we see that wave functions $P\in\cHQJ$ are $\SU(2)$-invariant
(vanish under the closure constraints) and are eigenvectors of the
$\tr\,\cC$-operator with eigenvalue $2J$.

Second, operators $\wM$ and $\widehat{(\tr\,\cC)}$
acting on the Hilbert space $\cHQJ$
(i.e., on $\SU(2)$-invariant
functions vanishing under the closure constraints) satisfy the same quadratic constraints as the
$\u(N)$-generators $E_{ij}$:
\begin{equation}
\widehat{(\tr\,\cC)}=\sum_k \wM_{kk},\qquad \sum_k \wM_{ik}\wM_{kj}
\,=\, \wM_{ij}\left(\f {\widehat{(\tr\,\cC)}}2 +N-2\right).
\end{equation}
This allows us to get the value of the (quadratic) $\U(N)$-Casimir
operator on the space $\cHQJ$:
$$
\sum_{ik}\wM_{ik}\wM_{ki}\,=\, \widehat{(\tr\,\cC)}\left(\f
{\widehat{(\tr\,\cC)}}2 +N-2\right) \,=\, 2J(J+N-2).
$$
Therefore, we can conclude that the quantization of our spinors and
$M$-variables exactly matches the  $\u(N)$-structure on the
intertwiner space (with the exact same ordering):
\begin{equation}
\cHQJ \sim\cHNJ,\quad \wM_{ij}\,=\,E_{ij},\quad
\widehat{(\tr\,\cC)}\,=\,E.
\end{equation}

In the third place, turning to the $\wQ_{ij}$-operators, it is straightforward to
check that they have the exact same action that the $F^\dag_{ij}$
operators, they satisfy the same Lie algebra commutators
\Ref{commEF} and the same quadratic constraints
(\ref{constraint1}-\ref{constraint3}). Clearly, the simple
multiplicative action of an operator $\wQ_{ij}$ sends a polynomial in
$\pP_J[Q_{ij}]$ to a polynomial in $\pP_{J+1}[Q_{ij}]$.
Reciprocally, the derivative action of $\wbQ_{ij}$ decreases the
degree of the polynomials and maps  $\pP_{J+1}[Q_{ij}]$ onto
$\pP_J[Q_{ij}]$.

Finally, let us consider the scalar product on the space of
polynomials $\pP[Q_{ij}]$. There seems to be a unique measure (up to a global factor)
compatible with the correct
Hermiticity relations for $\wM_{ij}$ and $\wQ_{ij},\wbQ_{ij}$. It is given by:
\begin{equation}
\label{scalarp} \forall \phi,\psi\in \pP[Q_{ij}],\quad
\la\phi|\psi\ra \,\equiv\, \int \prod_id^4z_i\,e^{-\sum_i \la
z_i|z_i\ra}\, \overline{\phi(Q_{ij})}\,\psi(Q_{ij})\,.
\end{equation}
Then it is easy to check that we have $\wM^\dag_{ij}=\wM_{ji}$ and
$\wQ^\dag_{ij}=\wbQ_{ij}$ as desired.

The spaces of homogeneous polynomials
$\pP_J[Q_{ij}]$ are  orthogonal with respect to this scalar product.
The quickest way to realize that is to consider the
operator $\widehat{(\tr\,\cC)}$, which is Hermitian with respect to
this scalar product and takes different values on spaces
$\pP_J[Q_{ij}]$ for different values of $J$. Thus these spaces
$\pP_J[Q_{ij}]$ are orthogonal to each other.

This concludes the quantization procedure, showing that the
intertwiner space for $N$ legs and fixed total area $J=\sum_i j_i$
can be constructed as the space of homogeneous polynomials in $Q_{ij}$
of degree $J$. We obtain a description of
intertwiners as anti-holomorphic wave-functions of spinors
$z_i$ constrained by the closure conditions\footnote{An
alternative construction \cite{Borja:2010rc},
which can be considered as ``dual'' to the representation defined
above, can also be carried out based on coherent states for the oscillators. This approach yields,
upon quantization, the framework of the $\U(N)$ coherent intertwiner states
introduced in \cite{Freidel:2010tt} and further developed in
\cite{Dupuis:2010iq}.}.


\section{Classical action and effective dynamics}

Once we have constructed the Hilbert space of a single intertwiner,
and in order to make contact with the standard spin network
formulation of loop quantum gravity, we want to apply this framework
to graphs. Therefore, we will have several intertwiners glued
together according to the corresponding graph structure. We will
then write an action principle in terms of spinors, compatible with
the Poisson bracket structure (\ref{bracketz}). In order to do so,
we need to take into account the closure constraints, but also the
matching conditions coming from the gluing of intertwiners.

The key element to elaborate the connection between the spinor
formalism and the standard formulation of LQG is the reconstruction
of the $\SU(2)$ group element $g_e$ (the holonomy) associated to an
edge in terms of the spinors \cite{Freidel:2010bw}. Let us consider
an edge $e$ with a spinor at each of its end-vertices
$z_{s(e),e}$ and $z_{t(e),e}$. There exists a unique $\SU(2)$ group
element mapping one onto the other. More precisely, the requirement
that
\begin{equation}
g_e\f{|z_{t(e),e}\ra}{\sqrt{\!\la
z_{t(e),e}|z_{t(e),e}\ra}}=\!\!\f{|z_{s(e),e}]}{\sqrt{\!\la
z_{s(e),e}|z_{s(e),e}\ra}},
\quad\!\! g_e\f{|z_{t(e),e}]}{\sqrt{\!\la
z_{t(e),e}|z_{t(e),e}\ra}}=-\f{|z_{s(e),e}\ra}{\sqrt{\!\la
z_{s(e),e}|z_{s(e),e}\ra}},
\quad\!\! g_e\in\SU(2),\nonumber
\end{equation}
i.e., that $g_e$ map the (normalized) source spinor
to the dual of the target spinor, uniquely fixes the value of $g_e$ to:
\begin{equation}
g_e\,\equiv\, \f{|z_{s(e),e}]\la z_{t(e),e}|-|z_{s(e),e}\ra
[z_{t(e),e}|} {\sqrt{\la z_{s(e),e}|z_{s(e),e}\ra\la
z_{t(e),e}|z_{t(e),e}\ra}}.
\end{equation}

\begin{figure}[h]
\begin{center}
\includegraphics[height=60mm]{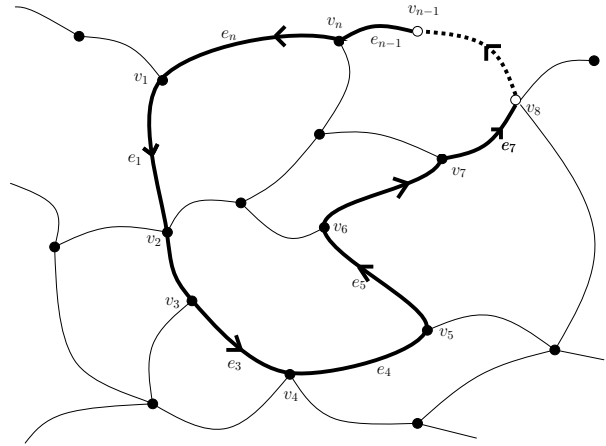}
\caption{The loop $\cL=\{e_1,e_2,..,e_n\}$ on the graph $\Gamma$.}
\label{loop_fig}
\end{center}
\end{figure}

The group elements  $g_e(z_{s(e),e},z_{t(e),e})\in\SU(2)$ commute
with the matching conditions, ensuring that the energy of the
oscillators on the edge $e$ is the same at both vertices $s(e)$ and
$t(e)$. However, they are obviously not invariant under $\SU(2)$
transformations. As we know from loop quantum gravity, in order to
construct $\SU(2)$-invariant observables, we need to consider the
trace of holonomies around closed loops, i.e., the oriented product
of group elements $g_e$ along closed loops $\cL$ on the graph:
\be
G_\cL\,\equiv\, \overrightarrow{\prod_{e\in\cL}}g_e.\nonumber
\ee
For the sake of simplicity, let us assume that all the edges of the
loop are oriented the same way, so we can number the edges
$e_1,e_2,..e_n$ with $v_1=t(e_n)=s(e_1)$, $v_2=t(e_1)=s(e_2)$, etc (see figure \ref{loop_fig}).
Then, we can explicitly write the
holonomy $G_\cL$ in terms of the spinors:
\be
\tr\, G_\cL \,=\,\tr g(e_1)\ldots g(e_n)=\tr \f{\prod_i
\left(|z_{v_i,e_i}]\la z_{v_{i+1},e_i}|-|z_{v_i,e_i}\ra
[z_{v_{i+1},e_i}|\right)}
{\prod_i\sqrt{\la z_{v_i,e_i}|z_{v_i,e_i}\ra\la
z_{v_{i+1},e_i}|z_{v_{i+1},e_i}\ra}}.\nonumber
\ee

Or, if instead of factorizing this expression by edges, we group
the terms by vertices, we obtain:
\be
\tr\, G_\cL \,=\, \sum_{r_i=0,1} (-1)^{\sum_ir_i} \f{\prod_i \la
\varsigma^{r_{i-1}} z_{v_i,e_{i-1}}\, |\, \varsigma^{1-r_i}
z_{v_i,e_i}\ra} {\prod_i \sqrt{\la z_{v_i,e_{i-1}}|
z_{v_i,e_{i-1}}\ra\la z_{v_i,e_i}| z_{v_i,e_i}\ra}},\nonumber
\ee
where the $\varsigma^{r_i}$ records whether we have the term
$|z_{v_i,e_i}]\la z_{v_{i+1},e_i}|$ or $|z_{v_{i},e_i}\ra
[z_{v_{i+1},e_i}|$ on the edge $e_i$, with $r_i=0,1$
(recall that $\varsigma$ is the anti-unitary map sending a spinor
$|z\ra$ to each dual $|z]$).

Depending on the specific values of the $r_i$ parameters, the scalar
products at the numerators are given by the matrix elements of $M^i$
or $Q^i$ at the vertex $i$. Since these matrices are by definition
$\SU(2)$-invariant (they commute with the closure conditions), this
is a consistency check, \emph{a posteriori}, that the holonomy $\tr\, G_\cL$
correctly provides a $\SU(2)$-invariant observable.

Taking into account the various possibilities for  the signs
$(-1)^{r_i}$, we can write the holonomy
\be
\tr \,G_\cL\,=\, \sum_{r_i=0,1} (-1)^{\sum_ir_i}
\f{\cM_\cL^{\{r_i\}}}{\prod_i \sqrt{\la z_{v_i,e_i}|
z_{v_{i+1},e_i}\ra}},
\ee
where the objects $\cM_\cL^{\{r_i\}}$ correspond to each term with a
fixed set $\{r_i\}$:
\beq
\cM_\cL^{\{r_i\}} &\equiv&\! \prod_i r_{i-1}r_i\bQ^i_{i,i-1}+
(1-r_{i-1})r_iM^i_{i-1,i}+ r_{i-1}(1-r_i)M^i_{i,i-1}+
(1-r_{i-1})(1-r_i)Q^i_{i,i-1} \nonumber\\ &=&\! \prod_i \la
\varsigma^{r_{i-1}} z_{v_i,e_{i-1}}\, |\, \varsigma^{1-r_i}
z_{v_i,e_i}\ra.\nonumber
\eeq

Using these ingredients, we are going to write an action principle
for this formalism. For this, we have to keep in mind the graph
structure, with intertwiners at the vertices, glued together along
edges. The phase space therefore consists of spinors $z_{v,e}$
(where $e$ are edges attached to the vertex $v$, i.e., such that
$v=s(e)$ or $v=t(e)$) which we constrain by the closure conditions
$\vcC^v$ at each vertex $v$ and the matching conditions on each edge
$e$. The corresponding action thus reads:
\begin{equation}
S_0^{\Gamma}[z_{v,e}] = \!\!\int \!\!dt\,\sum_v \sum_{e|v\in\pp
e}\!\!\! \left(-i\la z_{v,e}|\pp_t z_{v,e}\ra +\la
z_{v,e}|\Lambda_v| z_{v,e}\ra\right)\!
 +\sum_e \rho_e\!\left(\!
 \la z_{s(e),e}|z_{s(e),e}\ra\!\!-\!\!\la z_{t(e),e}|z_{t(e),e}\ra\!
 \right)\!,\nonumber
\end{equation}
where the $2\times 2$ Lagrange multipliers  $\Lambda_v$, satisfying
$\tr\,\Lambda_v=0$, impose the closure constraints and the Lagrange
multipliers $\rho_e\in\R$ impose the matching conditions. All the
constraints are first class, they generate $\SU(2)$ transformations
at each vertex and $\U(1)$ transformations on each edge $e$.

Analogously, this system can be parameterized in terms of $N_v\times
N_v$ unitary matrices $U^v$ and the parameters $\lambda_v$. The
matrix elements $U^v_{ef}$ refer to pairs of edges $e,f$ attached to
the vertex $v$. As mentioned before, the closure constraints are
automatically encoded in the requirement of unitarity for $U^v$.
It remains, then, to impose the matching conditions
$M^{s(e)}_{ee}-M^{t(e)}_{ee}=0$ on each edge $e$ (the matrices
$M^v=\lambda_v\,U^v\Delta (U^v)^{-1}$ being functions of both
$\lambda_v$ and $U^v$). So in this case, the action reads:
\begin{equation}
S_0^{\Gamma}[\lambda_v,U^v]\,=\, \int dt \,
\sum_v\left(-i\,\lambda_v \tr\,U^v\Delta \pp_t {U^v}^\dag \,-\,\tr
\Theta_v\,(U^v{U^v}^\dag-\id) \right) +\sum_e
\rho_e(M^{s(e)}_{ee}-M^{t(e)}_{ee}),\nonumber
\end{equation}
where the $\rho_e$ impose the matching conditions while the
$N_v\times N_v$ matrices $\Theta_v$ are the Lagrange multipliers for
the unitarity of the matrices $U^v$.

This free action describes the classical kinematics of spin networks
on the graph $\Gamma$. We can now add interaction terms to
this action. Such interaction terms should be built with generalized
holonomy observables $\cM_\cL^{\{r_i\}}$, thus trivially satisfying
the closure and matching conditions. A proposal for
a classical action for spin networks with non-trivial
dynamics was made in \cite{Borja:2010rc}:
\begin{equation}
S_{\gamma_\cL^{\{r_i\}}}^{\Gamma}=S_0^{\Gamma}\,+\, \int
dt\,\sum_{\cL,\{r_i\}}
\gamma_\cL^{\{r_i\}}\,\cM_\cL^{\{r_i\}},\nonumber
\end{equation}
where the $\gamma_\cL^{\{r_i\}}$ are the coupling constants giving
the relative weight of each generalized holonomy in the full
Hamiltonian. This action was then applied in particular to a specific model based
on a 2-vertex graph. We refer the reader to the original paper for more details on the
analysis of the dynamics derived from this Hamiltonian.


\section{Spinor representation and Holomorphic methods in LQG}

There is an interesting twist in the utility of this spinorial
framework for LQG \cite{Livine:2011gp}. These methods are extremely
powerful in order to understand the geometrical interpretation of
the spin network states. The single-edge Hilbert
space in LQG is given by
$\mathcal{H}_e:=L^2(\SU(2),dg)$, i.e., the space of square integrable
functions over the group $\SU(2)$. This Hilbert space can be
obtained as the quantization of a cotangent bundle over $\SU(2)$
acting as the classical phase space ($T^*\SU(2)\approx
\SU(2)\times \su(2)$). At this point, instead of the usual
coordinatization of this classical phase space given by $(g,X)$ --a
group and a Lie algebra element respectively--, a pair of
$\mathbb{C}^2$-spinors $(|z\rangle,
|\tilde{z}\rangle)$ recovering the classical physics in
$T^*\SU(2)$ can be employed.

As we have already seen before it is possible to obtain the
$\SU(2)$ group elements associated to such a pair of spinors. We are going
to review that discussion simplifying the notation and taking a slightly
different choice of mapping between the spinors and the group element and
its inverse\footnote{In order to remain faithful to the paper
by Livine and Tambornino \cite{Livine:2011gp}, we adopt in this
section the notation and the choice of the Poisson bracket structure
adopted by them. The only difference with respect the one presented
in section 4 is a sign in the Poisson bracket, thus switching from
an antiholomorphic to a holomorphic prescription.}. We start
with spinors $|z\rangle$ and $|\tilde{z}\rangle$, living
at the source and the target vertices of a given edge, then we can
write the corresponding $\SU(2)$-element and its inverse as follows:
\begin{eqnarray}
g(z,\tilde{z}) &=& \dfrac{ |z\rangle [\tilde{z}|- |z] \langle
\tilde{z}|}{\|z\|\,\|\tilde{z}\|}\,,\nn
g^{-1}(z,\tilde{z})&=&\dfrac{-|\tilde{z}\rangle [z|+
|\tilde{z}]\langle z|}{\|z\|\,\|\tilde{z}\|}\,.\nonumber
\end{eqnarray}
It is straightforward to show that, under a local
$\SU(2)$-transformation $h\in \SU(2)$ in the defining representation
of the group, $g(z,\tilde{z})$ transforms as the holonomy of a
$\SU(2)$-connection:
\be
g(z,\tilde{z}) \longmapsto h_1 g(z,\tilde{z})h_2^{-1}\,.\nonumber
\ee
Moreover, each spinor is mapped to $\mathbb{R}^3$ (up to a phase)
yielding area vectors corresponding to the different faces of
an elementary polyhedron. In this sense, the Hilbert space for a
graph with $E$ edges can be regarded as the space of glued polyhedra
covering a piecewise flat manifold (this is ensured by the
imposition of the proper matching conditions). Following this line of
thought, it is easy to show that the $\su(2)$-elements
$X(z):=\vec{X}(z)\cdot \vec{\sigma}$ and
$\tilde{X}(\tilde{z}):=\vec{\tilde{X}}(\tilde{z})\cdot
\vec{\sigma}$, associated with the normals to the glued faces of two
polyhedra, are related through the expression:
$$\tilde{X}=g^{-1}Xg.$$

\medskip

On the other hand, we can endow the space
$\mathbb{C}^2\times\mathbb{C}^2$ with a symplectic structure given
by
$$\{\bar{z}^i,z^j\}=i\delta_i^j=\{\bar{\tilde{z}}^i,\tilde{z}^j\}\,,$$
thus obtaining a proper phase space. This phase space has to be
equipped with a constraint forcing the two
spinors to have the same length (matching condition). This constraint
is given by $\|z^2\|-\|\tilde{z}^2\|$ and it can be noticed that it
generates $\U(1)$ transformations. Nevertheless, the group and
Lie-algebra elements are invariant under these transformations by
construction. The interesting point here is that, using this symmetry, one can arrive to
the cotangent bundle of $\SU(2)$ by means of a $\U(1)$-gauge
reduction:
\be
\mathbb{C}^2\times\mathbb{C}^2/\!\!/\U(1)/\mathbb{Z}_2\simeq
T^*\SU(2)\setminus \{|X|=0\}\,,\nonumber
\ee
where $/\!\!/$ denotes a double quotient (see \cite{Livine:2011gp}
for details). This space has indeed the correct symplectic structure
--that arises in terms of $g$ and $X$-- written in terms of spinors.  In
this sense, given an edge of a graph, one can either assign it a
pair $(g,X)$ or a pair of spinors $(|z\rangle$,
$|\tilde{z}\rangle)$. This way, the degrees of freedom
are shifted from the edge to its vertices.

\medskip

Once we are convinced that this correspondence between spinors and
the $\SU(2)$-group and algebra elements works, we can also find
the expression of the Haar measure $dg$ in terms of spinors. The
result is really simple, the Haar measure is just the product of two
Gaussian measures
\be
\int_{\SU(2)}dg=\int_{\mathbb{C}^2\times\mathbb{C}^2}d\mu(z)d\mu(\tilde{z}).\nonumber
\ee
In order to arrive to this result one has to be careful with the
redundancies introduced by the spinors ($8$ degrees of freedom
instead of the $3$ in a group element $g$) which means that one
has to make use of a \emph{twisted rotation} (as defined in \cite{Livine:2011gp})
in order to implement a
$\SU(2)$-transformation leaving the group element invariant. Taking
this into account the former relation can be proven by computing the
scalar product between two representation matrices of $\SU(2)$ in
terms of spinors. We refer the interested reader to the original paper
for more details.

\medskip

The procedure presented here provides an alternative quantization
for $T^*\SU(2)$, where the Hilbert space for an edge $e$,
$\mathcal{H}^{spin}_e$ is obtained after an appropriate
gauge-reduction, implementing the proper matching conditions on that
edge, of the Bargmann space of holomorphic square-integrable
functions on both spinors. This is, in some sense, an opposite way of
thinking with respect to the usual picture, where the group and Lie
algebra elements are taken as the fundamental variables. Here, the spinorial
variables are considered as fundamental and the group
and Lie algebra elements arise as composite objects. We can also
build ladder operators in terms of spinors and derive the
standard holonomy and flux-operators of the theory.

We are considering two quantizations of the same classical phase space,
resulting from the use of two different polarizations, and we arrive at the usual
LQG Hilbert space $\mathcal{H}_e$ with the standard
variables or to $\mathcal{H}^{spin}_e$ in terms of spinors. The relevant
question then is, are those Hilbert spaces unitarily equivalent? The answer is in the
affirmative. Indeed, one can construct a unitary map between those
Hilbert spaces
\be
\mathcal{T}:\mathcal{H}_e\rightarrow \mathcal{H}_e^{spin}\nonumber
\ee
employing a method based on a modification of the Segal-Bargmann
transform. This map sends $SU(2)$ representation matrices
to holomorphic functions in $(|z\rangle , |\tilde{z}\rangle)$ and it can be regarded
as the restriction to
the holomorphic part of the group element written in
terms of spinors (see \cite{Livine:2011gp} for details). Furthermore, this map can be straightforwardly generalized to an
arbitrary graph and can be shown to be compatible with the
inductive limit construction used to define the Hilbert
space of the continuum theory.
Therefore, there is a strong formal consistency
for this reformulation of LQG, which captures the same physics while
opening new ways to tackle the open problems in the theory,
such as the lack of a well defined semiclassical limit or the
geometric interpretation of the spin networks states.


\section{Conclusions}

We have reviewed here the $\U(N)$ and spinorial framework for LQG.
This new point of view has become a very useful tool to
deal with certain fundamental problems in LQG. In particular, it was
interesting for the study of a simple model (the $2$-vertex model)
with a pair of nodes and an arbitrary number of links
\cite{Borja:2010gn,Borja:2010rc,Borja:2010hn}. In these papers,
a plausible dynamics was proposed for this system (at the
quantum and classical level) and some striking mathematical
analogies with loop quantum cosmology were explored. Also, using
this framework, it was possible to study semiclassical states
\cite{Freidel:2010tt} and the simplicity constraints
\cite{Dupuis:2010iq}.

Let us revisit the main points presented in this review:
\begin{itemize}
\item Making use of the Schwinger representation, it is possible to
write $\SU(2)$ invariant operators acting on the Hilbert space of
intertwiners in LQG \cite{Girelli:2005ii}.

\item A key point for the $\U(N)$ construction is that the LQG
Hilbert space of intertwiners with $N$-legs and fixed total area $J$
carries an irreducible $\U(N)$ representation
\cite{Freidel:2009ck}.

\item One can show that the full space of $N$-leg intertwiners
is endowed with a Fock space structure, with annihilation and
creation operators given by the operators $F_{ij}$ and
$F_{ij}^\dagger$.

\item We have presented a classical framework based on spinors whose
quantization (using holomorphic methods) leads back to the $\U(N)$
framework for the Hilbert space of intertwiners in LQG
\cite{Borja:2010rc}.

\item An action principle for the spinor phase
space has been described. Then, using the relation given in \cite{Freidel:2010bw}
between spinors and $\SU(2)$ group elements, an
expression for the holonomy operators was built. Taking advantage of these
$\SU(2)$ and $\U(1)$ invariant operators, it was possible to give a
generic interaction term (whose quantization is direct) for this
framework.

\item Finally, we gave a glimpse of the formal and rigorous construction of the
spinor representation and holomorphic quantization techniques for
the whole LQG kinematical Hilbert space, as it was developed by
Livine and Tambornino \cite{Livine:2011gp}. There, it was shown that
although the choice of the spinors as fundamental variables of the
theory implies a different choice of polarization (and in general
different polarizations lead to inequivalent quantizations), in this
case it is possible to construct explicitly a unitary map
$\mathcal{T}$ between the usual LQG Hilbert space and the one derived in
terms of the spinors.

\end{itemize}

Both the $\U(N)$ framework and the spinor representation are
promising ways to deal with the most important open problems within
LQG, namely, the semiclassical states and the dynamics. They
may also become useful computational tools to calculate physical
quantities within the theory, such as correlation functions.
Besides, this new point of view may be useful in the context of
group field theory. The possibility of having Feynman amplitudes
expressed as integrals over spinor variables or the study of the
renormalization properties of group field theories are interesting
problems where the spinor representation can play a crucial role.

To conclude, we would like to remark that this framework
certainly constitutes a viewpoint that opens new ways of research,
both in loop quantum gravity and in group field theory and spin
foams, that are worth exploring.


\subsection*{Acknowledgements}

We would like to thank Etera R. Livine for endless discussions.

This work was in part supported by the Spanish MICINN research
grants FIS2008-01980, FIS2009-11893 and ESP2007-66542-C04-01 and by the grant NSF-PHY-0968871.
IG is supported by the Department of Education of the Basque Government under the
``Formaci\'{o}n de Investigadores'' program.


\end{document}